\documentclass{PoS}
\usepackage{amssymb}
\usepackage{bm}
\usepackage{latexsym}

\newcommand{\case}[2]{\ensuremath{{\textstyle\frac{#1}{#2}}}}
\newcommand{\Chi}{\ensuremath{X}}

\newcommand{\Dirac}{\ensuremath{{D\kern -0.65em /}}}

\newcommand{\half}{\ensuremath{{\textstyle\frac{1}{2}}}}

\newcommand{\tr}{\mathop{\mathrm{tr}}}

\newcommand{\Creutz}{\cite{Creutz:2006ys,Creutz:2006wv,Creutz:2007yg,Creutz:2007nv,Creutz:2007rk}}

\title{Lattice gauge theory with staggered fermions: \\
how, where, and why (not)} % {\thanks{Invited plenary talk.}}

\ShortTitle{Staggered fermions: why not!}

\author{\speaker{Andreas S. Kronfeld} \\
		Theoretical Physics Department,
		Fermi National Accelerator Laboratory,$\!$%
		\thanks{Operated by Fermi Research Alliance, LLC, under Contract
		No.~DE-AC02-07CH11359 with the United States Department of Energy.}~
		Batavia, Illinois, USA \\
		E-mail: \email{ask@fnal.gov}}

\abstract{Many results from lattice QCD of broad importance to particle
and nuclear physics are obtained with 2+1 flavors of staggered sea
quarks.
In the continuum limit, staggered fermions yield four species, called
tastes.
To reduce the number of tastes to one (per flavor), the simulation
employs the fourth root of the four-taste staggered fermion determinant.
This talk surveys evidence in favor of this procedure, refutes recent
criticisms, and reviews recent algorithmic and technical improvements.
Physics results are covered in other plenary talks.}

\FullConference{The XXV International Symposium on Lattice 
Field Theory \\ July 30--4 August 2007 \\ Regensburg, Germany}

\begin{document}

\section{Why this paper?}

\vspace*{-0.1em}
Lately lattice QCD has enjoyed considerable success.
In 2003 the HPQCD, MILC, and Fermilab Lattice Collaborations found that
calculations based on Monte Carlo simulations agreed with experiment at
the 1--3\% level for a wide variety of physical
quantities~\cite{Davies:2003ik}.
During the next two years the same techniques were used to predict the
shape of the $D\to Kl\nu$ semileptonic form factor~\cite{Aubin:2004ej},
the mass of the $B_c$ meson~\cite{Allison:2004be},
and the decay constants of the $D$ and $D_s$ mesons~\cite{Aubin:2005ar}.
Each of these results was subsequently confirmed by experimental 
measurements~\cite{Kronfeld:2005fy}.
Lattice QCD was also used to determine the strong
coupling~$\alpha_s$~\cite{Mason:2005zx}.
These results have made a strong impression in the fields of particle
physics and nuclear physics, and a long-standing claim of lattice gauge
theorists is now generally accepted: 
the combination of numerical simulation~\cite{Creutz:1980wj} and chiral
perturbation theory~\cite{Bijnens:2007yd} is a sound way to solve gauge
theories, such as~QCD.

The key to this development was the incorporation of 2+1 flavors of sea
quarks with the \emph{fastest} technique, ``improved staggered fermions
with the Asqtad action'' \cite{Lepage:1998vj}.
Unsurprisingly for computational science, the fastest technique has some 
unresolved theoretical questions hanging over it.
This state of affairs presents a great opportunity.
It is simple common sense to reproduce the results of
Refs.~\cite{Davies:2003ik,Aubin:2004ej,Allison:2004be,Aubin:2005ar,%
Mason:2005zx} with theoretically cleaner treatments of sea quarks.
Such methods require more computation, so one can argue (persuasively)
to devote more computer resources to lattice QCD than in the past.
At the same time, there is a strong motivation (and obligation) to 
understand theoretically how, where, and why staggered fermions work 
so well.

The crux of the unresolved issues is the fermion doubling problem.
Staggered fermions cope with the problem partially, reducing the number 
of species from sixteen (in four space-time dimensions) to four.
The empirically successful results cited above all use ensembles 
generated and made publicly available by the MILC 
Collaboration~\cite{Bernard:2001av}.
To reduce the number of fermion species (per flavor) from four to one,
gauge fields in these ensembles have the partition
function~\cite{Hamber:1983kx}
\begin{equation}
	Z=\int\mathcal{D}U\;
		\prod_q\left[{\det}_4\left(\Dirac_{\rm stag}+m_q\right) 
		\right]^{1/4} e^{-S_{\rm gauge}},
	\label{eq:rooting}
\end{equation}
where $S_{\rm gauge}$ is the action for the gluons, $\Dirac_{\rm stag}$
is the (improved) staggered discretization of the Dirac operator, and
the subscript 4 on the determinant emphasizes the underlying number of
species.
The question, then, is whether 
$[\det_4(\Dirac_{\rm stag}+m_q)]^{1/4}$ is legitimate.

This so-called ``rooting procedure'' is controversial.
% Several facets of the controversy might interest a sociologist of 
% science, but most of them lie beyond what I was asked to talk about.
%
% criticisms are not written up
% criticisms are not based on careful reading of long papers
% attacks are often personal
% no controversy about Taiwan work (quenched + linear chiral extrap)
% no controversy about rooting in thermodynamics
% 
Critics seem to accept that Eq.~(\ref{eq:rooting}) is valid in
perturbation theory, where internal fermion loops are simply multiplied
by~$\case{1}{4}$.
Several interested and disinterested parties are investigating 
non-perturbative aspects, sometimes finding undesirable features though 
not, to my knowledge, fatal flaws.
An apparent exception is the work of Michael 
Creutz~\cite{Creutz:2006ys,Creutz:2006wv,Creutz:2007yg,Creutz:2007nv}, 
including these Proceedings~\cite{Creutz:2007rk}, which does claim that 
rooting fails.
Although these papers have already been refuted~\cite{Bernard:2006vv},
the organizers of Lattice 2007 invited me to respond to Creutz's claims
``on behalf of the staggered community.''  My charge is to ``increase
the confidence of the \emph{wider} lattice community in the rooting
procedure'' (\emph{italics} added).

The organizers also asked me to comment on the relevance of lattice 
gauge theory to CERN's Large Hadron Collider (LHC).
This request and the need to anticipate new critiques have shaped 
the organization of this talk.
Section~\ref{sec:community} is an essay presenting the view that the LHC
era will require fast lattice calculations and, thus, a broader
understanding of validity of the rooting  procedure.
Sections~\ref{sec:staggered} and \ref{sec:rooting} disentangle the main 
issues. \pagebreak
Even without rooting, staggered fermions bring in complications, which
are reviewed in sect.~\ref{sec:staggered}, focusing on a new quantum
number ``taste'' that characterizes the remaining species doubling.
Furthermore, even with full SU($4n_f$) taste symmetry, rooting has its
peculiarities, which may be unfamiliar but should not be controversial;
they are explained in sect.~\ref{sec:rooting}.
These two strands are then brought together in sect.~\ref{sec:synthesis}
to discuss rooted staggered fermions.
It is then possible to refute Refs.~\Creutz\ relatively tersely, in 
sect.~\ref{sec:anti_Creutz}.
Section~\ref{sec:new_developments} covers some new developments, and 
sect.~\ref{sec:conclusions} offers a summary and perspective.

\section{The lattice community in the LHC era}
\label{sec:community}

To represent the ``staggered community'' before the ``wider lattice 
community'' it helps to think about who belongs to each of them.
The wider lattice community simply consists of those who attend
conferences on lattice field theory and submit e-prints to the hep-lat
section of the arXiv.
The staggered community includes not only those who generate large 
ensembles of SU(3) gauge fields with rooted, staggered sea quarks, 
but also those (like me) who use them.
The staggered community also reaches beyond hep-lat.
Many physicists without much expertise in lattice gauge theory have a
stake in the validity of the rooting procedure.
For example, the Heavy Flavor Averaging Group~\cite{Barberio:2007cr} and
similar enterprises use lattice QCD results to help them understand
whether non-Standard phenomena contribute to $B$-meson decays.
It is therefore useful to take the staggered community to be everyone
who thinks the validity of staggered fermions is worthy of study.

The organizers' charge implies that there may be some in the wider
lattice community who do not see themselves as part of the staggered
community, even as defined here.
That would be unfortunate.
Despite the ``controversial'' nature of Eq.~(\ref{eq:rooting}), the
arguments against it are difficult to pin down, because they are not
documented in the scientific literature.
Mike Creutz is to be commended for writing up his critique, making it 
possible to decide whether his qualms are correct.
On the other hand, some members of the lattice community seem to prefer 
sniping from the sidelines.
They may hope that increases in computer resources will make the
controversy moot, by allowing other methods for sea quarks to catch up.
The status of other methods is surveyed in Refs.~\cite{others}, and
one may judge for oneself.
My reading is that other methods have not caught up, partly because 
algorithms for the staggered sea are not standing 
still~\cite{Clark:2006wp}.

Future calculations are more salient than past efforts, so it is worth
looking ahead to the LHC era.
The focus of the LHC's physics program will be on the 
terascale,%~\cite{joe},
\footnote{Using the term ``terascale'' for the teraelectronvolt energy
scales was inspired by terascale (\emph{i.e.}, teraflop/s and terabyte)
computing (J.D. Lykken, private communication).}
where we expect to find the agents of electroweak symmetry breaking.
They may be strongly coupled and, if so, the need for numerical lattice
gauge theory will skyrocket.
A~lot will be at stake, and the fastest way to elucidate the physics
will prevail.
Even if physics at the terascale is weakly-coupled or, worse, boring, 
a need will remain for precision and accuracy in $f_K$, $f_B\sqrt{B_B}$, 
\emph{etc}.
In addition, searches for non-Standard particles will profit from 
good calculations of moments of parton densities, especially the gluon 
density, so that signal and background cross sections can be 
calculated.

In summary, whether for QCD or for other gauge theories, the fastest
correct technique for simulating the fermion sea will remain a key tool
during the LHC era.
New researchers, young and old, may turn to lattice gauge theory, and 
they will expect that we established experts have \pagebreak understood whether 
rooted staggered fermions are valid, or not.
For this reason I think the whole lattice community should engage in an
open-minded scientific debate, and skeptics should submit their
criticisms of rooted staggered fermions to the hep-lat~arXiv.

\section{Staggered fermions without rooting}
\label{sec:staggered}

Most of the complexity of rooted staggered quarks has nothing to 
do with the rooting procedure.
Instead, it stems from the way spacetime and flavor-like symmetries
of four (or $4n_f$) Dirac fermions emerge in the continuum limit.
The aim of this section, therefore, is to give a brief review of the
definitions and symmetries of staggered fermions and their connection
with the continuum limit.
Most of this is not rigorously \emph{proven}, but it is fair to say that
it is \emph{established}, principally because numerical simulations
shore up the theoretical picture.

The simplest discretization of fermions replaces Dirac's covariant 
derivative with a nearest-neighbor interaction.
The resulting ``na\"ive'' action is
\begin{equation}
	S_{\textrm{\scriptsize na\"ive}} = 
		\case{1}{2} a^3 \sum_{x,\mu} \bar{\Upsilon}(x)\gamma_\mu\left[
		U_\mu(x)\Upsilon(x+\hat{\mu}a) - 
		U^\dagger_\mu(x-\hat{\mu}a)\Upsilon(x-\hat{\mu}a) \right] + 
		m_0a^4 \sum_x \bar{\Upsilon}(x)\Upsilon(x),
	\label{eq:naive}
\end{equation}
where each site $x$ possesses Grassmann variables $\Upsilon_\alpha^i$ and
$\bar\Upsilon_\alpha^i$, with $i$ and $\alpha$ the color and spinor 
indices.
This action is invariant under color SU($N_c$), lattice translations, and 
hypercubic rotations.
With $n_f$ flavors there is a ${\rm U_V}(n_f)\times{\rm U_A}(n_f)$
chiral flavor symmetry, softly broken by the masses.

The na\"ive action also possesses a remarkable SU(4) ``doubling symmetry''
\cite{Karsten:1980wd}, with fifteen $x$-dependent (anti-Hermitian) 
generators $B^A(x)$:
\begin{equation}
	B_\mu(x) = \gamma_\mu\gamma_5(-1)^{n_\mu}, \quad
	B_5(x) = i\gamma_5 \varepsilon(x), \quad
	B_\mu(x) B_5(x), \quad
	B_\mu(x) B_\nu(x)~~(\mu<\nu),
\end{equation}
where $n=x/a$, and $\varepsilon(x)=(-1)^{n_1+n_2+n_3+n_4}$.
The na\"ive fermion field transforms as
\begin{equation}
	\Upsilon(x) \mapsto e^{\omega^AB^A(x)}\Upsilon(x), \quad 
	\bar{\Upsilon}(x) \mapsto \bar{\Upsilon}(x)e^{-\omega^AB^A(x)}.
	\label{eq:doubling}
\end{equation}
The physical interpretation becomes clear in momentum space.
Consider, for example, a doubling transformation generated by $B_\mu$:
$\Upsilon(p) \mapsto \cos\omega\,\Upsilon(p) + 
\sin\omega\,\gamma_\mu\gamma_5\Upsilon(p+\hat{\mu}\pi/a)$.
In general, the doubling symmetries relate all 16 corners $\pi^A/a$ of the 
Brillouin zone, up to a shuffling of the Dirac index.
(See Eqs.~(\ref{eq:I-taste})--(\ref{eq:P-taste}), below, for a complete
list of the 4-vectors~$\pi^A$.)

The physical consequence of the doubling symmetry (in four dimensions) 
is that a single na\"ive fermion field~$\Upsilon$ corresponds to 16 
species of fermion.
The extra species are evident in vacuum polarization, leading to
$\beta_0=\case{11}{3}N_c-\case{2}{3}16n_f$ in the running of the gauge
coupling~\cite{Karsten:1980wd}.
The axial anomaly receives contributions from all 16 species, in the 
pattern~\cite{Karsten:1980wd}
\begin{equation}
	\sum_A \mathfrak{A}(\pi^A) = \left[
	\stackrel{(0,0,0,0)}{1} -
	\stackrel{(\pi,0,0,0)}{4} +
	\stackrel{(\pi,\pi,0,0)}{6} -
	\stackrel{(\pi,\pi,\pi,0)}{4} +
	\stackrel{(\pi,\pi,\pi,\pi)}{1} \right]\mathfrak{A} = 0,
	\label{eq:anomaly}
\end{equation}
where a typical $\pi^A$ hovers over the integer multiplicity of the 
species from that kind of corner.
The total anomaly vanishes because with the na\"ive action the 
flavor-singlet axial symmetry ${\rm U_A}(1)$ is exact.
Because the anomaly is wrong, na\"ive fermions do not seem to be what
one wants for~QCD.

The doubling symmetries can be rendered $x$-independent via a change of 
variables~\cite{Kawamoto:1981hw}:
\begin{equation}
	\Upsilon(x) = \Omega(x)\Chi(x), \quad
	\bar{\Upsilon}(x) = \bar{\Chi}(x)\Omega^{-1}(x), \quad
	\Omega(x)=\gamma_1^{n_1}\gamma_2^{n_2}\gamma_3^{n_3}\gamma_4^{n_4}.
	\label{eq:KawamotoSmit}
\end{equation}
Rewriting the na\"ive action in the new fermion fields $\Chi^i_\alpha$, 
$\bar\Chi^i_\alpha$, one finds
\begin{equation}
	S_{\textrm{\scriptsize na\"ive}} = 
		\case{1}{2} a^3 \sum_{x,\mu} \bar{\Chi}(x)\eta_\mu(x)\left[
		U_\mu(x)\Chi(x+\hat{\mu}a) - 
		U^\dagger_\mu(x-\hat{\mu}a)\Chi(x-\hat{\mu}a) \right] + 
		m_0a^4 \sum_x \bar{\Chi}(x)\Chi(x),
	\label{eq:naive-staggered}
\end{equation}
where the signs~$\eta_\mu(x)$ have replaced the Dirac matrices via
\begin{equation}
	\Omega^{-1}(x)\gamma_\mu\Omega(x\pm\hat{\mu}a) = 
		(-1)^{\sum_{\rho<\mu}n_\rho} =:  \eta_\mu(x).
	\label{eq:eta}
\end{equation}
Equation~(\ref{eq:naive-staggered}) can also be obtained by 
diagonalizing a maximal commuting subgroup of the doubling 
symmetry~\cite{Sharatchandra:1981si}.

In Eq.~(\ref{eq:naive-staggered}) the transformed spinor index is
sterile, so the number of degrees of freedom can be reduced four-fold,
yielding
\begin{equation}
	S_{\mathrm{stag}} = 
		\case{1}{2} a^3 \sum_{x,\mu} \bar{\chi}(x)\eta_\mu(x)\left[
		U_\mu(x)\chi(x+\hat{\mu}a) - 
		U^\dagger_\mu(x-\hat{\mu}a)\chi(x-\hat{\mu}a) \right] + 
		m_0a^4 \sum_x \bar{\chi}(x)\chi(x),
	\label{eq:staggered}
\end{equation}
where $\chi^i$ is a fermion field \emph{without} a spinor index.
A Hamiltonian formalism with one-component fermions and sign factors
instead of Dirac spinors and matrices was introduced by
Susskind~\cite{Susskind:1976jm}, extending work in 1+1 dimensions
by Banks, Kogut, and Susskind~\cite{Banks:1975gq}.
Because of the ubiquitous factors $(-1)^{n_\mu}$ the Euclidean formulation 
with $S_{\mathrm{stag}}$ and $\chi^i$ is called ``staggered fermions.''

The projection from the big $\Chi^i_\alpha$ to the little $\chi^i$ removes 
the SU(4) doubling symmetry, albeit in a not-so-straightforward way.
All other symmetries---except color and the vector flavor symmetries%
---become $x$-dependent \cite{Susskind:1976jm,Golterman:1984cy}.
For example, consider translations
\begin{equation}
	t_\mu: \left\{ %}
	\begin{array}{lcl}
		\Upsilon(x) \mapsto \Upsilon(x+\hat{\mu}a) 
		& \Rightarrow & 
		\Chi(x) \mapsto \zeta_\mu(x) \gamma_\mu \Chi(x+\hat{\mu}a) \\
		\bar{\Upsilon}(x) \mapsto \bar{\Upsilon}(x+\hat{\mu}a) 
		& \Rightarrow & 
		\bar{\Chi}(x) \mapsto \zeta_\mu(x) \bar{\Chi}(x+\hat{\mu}a)
			\gamma_\mu 
	\end{array} \right.,
	\label{eq:translations} 
\end{equation}
where
\begin{equation}
	\zeta_\mu(x) = \Omega^{-1}(x)\Omega(x\pm\hat{\mu}a)\gamma_\mu = 
		(-1)^{\sum_{\sigma>\mu}n_\sigma}.
	\label{eq:zeta}
\end{equation}
Because of the Dirac matrix in~(\ref{eq:translations}), we see that 
the projection from $\Chi^i_\alpha$ to $\chi^i$ does not commute with 
lattice translational symmetry.
On the other hand, a certain combination of translations and 
$B$~transformations called \emph{shifts} does survive the projection:
\begin{equation}
	S_\mu: \left\{ %}
	\begin{array}{lcl}
		\Upsilon(x) \mapsto -iB_\mu B_5\Upsilon(x+\hat{\mu}a) 
		& \Rightarrow & 
		\chi(x) \mapsto \zeta_\mu(x) \chi(x+\hat{\mu}a) \\
		\bar{\Upsilon}(x) \mapsto -i\bar{\Upsilon}(x+\hat{\mu}a)B_\mu B_5 
		& \Rightarrow & 
		\bar{\chi}(x) \mapsto \zeta_\mu(x) \bar{\chi}(x+\hat{\mu}a)\\
		U_\nu(x) \mapsto  U_\nu(x+\hat{\mu}a), & \forall \nu & 
	\end{array} \right.,
	\label{eq:shifts} 
\end{equation}
and $\Chi_\alpha$ transforms just like~$\chi$.
Similarly, rotations, spatial inversion, and charge conjugation become 
entangled with this residue of the doubling 
symmetries~\cite{Golterman:1984cy,Golterman:1984dn,Golterman:1985dz,%
Golterman:1986jf,Kilcup:1986dg,Joos:1987bp}.

Here we shall focus on issues related to the species content of staggered
fermions.
Acting on the fermion fields, the shifts anti-commute,
$S_\mu S_\nu=-S_\nu S_\mu$ ($\nu\neq\mu$), from which it follows that
the residue of the doubling symmetry is a discrete Clifford
group~$\Gamma_4$.
% Note to self:
% The projection does not remove the ``obvious'' $\Gamma_4$ subgroup of 
% $\mathbb{Z}_N^4\times$SU(4), but one with non-trivial translations 
% involved, leaving shifts as a good symmetry.
Because shifts translate the gauge field, a single $\Gamma_4$ symmetry 
arises for any~$n_f$.
Shift symmetry has two kinds of irreducible representations (irreps)---%
\emph{fermionic} and \emph{bosonic}---with representation matrices, 
respectively, 
\begin{eqnarray}
	\mathfrak{D}^{(-)}(S_\mu) & = & \xi_\mu e^{ip_\mu a},
	\label{eq:fermionic-irrep} \\ 
	\mathfrak{D}^{(+)}(S_\mu) & = & s_\mu^A e^{ip_\mu a},
	\label{eq:bosonic-irrep} 
\end{eqnarray}
where the physical momentum ranges over $p_\mu\in(-\pi/2a,% [
\pi/2a]$. %)
The fermionic representation is 4-dimensional, labeled by an index 
$t=1,2,3,4$, and the $\xi_\mu$ are $4\times 4$ matrices obeying the
Clifford algebra, $\{\xi_\mu,\xi_\nu\}=2\delta_{\mu\nu}$.
The sixteen 1-dimensional bosonic irreps are labeled by the corners~$A$ of 
the Brillouin zone and the pre-factors are signs, 
$s_\mu^A=e^{i\pi_\mu^A}$.
The index~$t$ and the label~$A$ denote a quantum number that, nowadays, is 
called \emph{taste}.

A colored particle's taste, just like its momentum, is made complicated
by SU($N_c$) color gauge symmetry.
It is enough, however, to consider color singlets, such as the 
mixed-action bilinear, which arises, for example, in heavy-light 
physics~\cite{Wingate:2002fh}.
Let
\begin{equation}
	H_t^{(\Gamma)}(x) = \bar{\Psi}_\alpha(x) \Gamma_{\alpha\beta}
		\Omega_{\beta t}(x) \chi(x),
	\label{eq:mixed-bilinear}
\end{equation}
where the matrix $\Omega$ reappears, now with the first index interpreted 
as a Dirac index, but the second index interpreted as taste 
index~\cite{Gliozzi:1982ib,KlubergStern:1983dg}.
Here the field $\bar{\Psi}$ represents a Wilson or Ginsparg-Wilson 
anti-fermion, so its spinor index has the conventional meaning.
Under shift symmetry (as one can easily verify), the meson field 
$H_t^{(\Gamma)}$ transforms as
\begin{equation}
	S_\mu: H_t^{(\Gamma)}(x) \mapsto 
		[\xi_\mu]_{tt'} H_{t'}^{(\Gamma)}(x+\hat{\mu}a), \quad
	\xi_\mu = \gamma_\mu^T ,
	\label{eq:mixed-translated}
\end{equation}
so Eq.~(\ref{eq:mixed-translated}) gives an explicit realization of 
Eq.~(\ref{eq:fermionic-irrep}).
Under rotations and spatial inversion $H_t^{(\Gamma)}$ transforms as a
scalar, vector, or tensor, \emph{etc.}, dictated by $\Gamma$, up to a
change of taste.
Baryons with three staggered quarks also have fermionic 
taste~\cite{Golterman:1984dn,Bailey:2006zn}.

Staggered-staggered mesons have bosonic taste~$A$.
Rotations imply degeneracies in an almost obvious way,%
\footnote{Strictly speaking, the degeneracies are among irreps of the
symmetry group of the transfer matrix, which entails cubic rotations,
not hypercubic rotations~\cite{Golterman:1986jf,Kilcup:1986dg}.
Thus, states with $\pi^A_4=0,\pi$ need not be degenerate; hence the
semicolons in Eqs.~(\ref{eq:V-taste})--(\ref{eq:A-taste}).
In practice they turn out to be nearly degenerate~\cite{Bernard:2001av}.}
leading to multiplets
\begin{eqnarray}
	I&: & \pi^A \in \{(0,0,0,0)\}, \label{eq:I-taste} \\
	V&: & \pi^A \in \{(\pi,0,0,0), (0,\pi,0,0), (0,0,\pi,0);
		(0,0,0,\pi)\}, \label{eq:V-taste} \\
	T&: & \pi^A \in \{(\pi,\pi,0,0), (\pi,0,\pi,0), (0,\pi,\pi,0); 
		(\pi,0,0,\pi), (0,\pi,0,\pi), (0,0,\pi,\pi)\}, \\
	A&: & \pi^A \in \{(0,\pi,\pi,\pi), (\pi,0,\pi,\pi), (\pi,\pi,0,\pi); 
		(\pi,\pi,\pi,0)\}, \label{eq:A-taste} \\
	P&: & \pi^A \in \{(\pi,\pi,\pi,\pi)\}.
	\label{eq:P-taste}
\end{eqnarray}
These multiplets are called the taste-singlet~$I$, the vector taste~$V$, 
the tensor taste~$T$, the axial vector taste~$A$, and the pseudoscalar 
taste~$P$.
Another hadron with bosonic taste is a heavy-light baryon consisting of 
two staggered light quarks and a Fermilab or NRQCD heavy 
quark~\cite{Gottlieb:2007ay}.

The staggered action, Eq.~(\ref{eq:staggered}), inherits softly broken
but otherwise exact chiral symmetries,
\begin{equation}
	\chi(x)       \mapsto e^{\theta^aT^a\varepsilon(x)} \chi(x), \quad
	\bar{\chi}(x) \mapsto \bar{\chi}(x) e^{\theta^aT^a\varepsilon(x)}, 
	\label{eq:axial}
\end{equation}
where ${T^a}^\dagger=-T^a$.
With $n_f$ flavors, this is simply the ${\rm U_A}(n_f)$ symmetry
manifest in Eq.~(\ref{eq:naive}), made $x$-dependent via
Eqs.~(\ref{eq:KawamotoSmit}).
Thus, one has axial currents and pseudoscalar densities
\begin{eqnarray}
	A^{\nu a}_P(x) & = & \half \eta^\nu(x) \varepsilon(x) \left[ 
		\bar{\chi}(x+\hat{\nu}a) U^\dagger_\nu(x) T^a \chi(x) - 
		\bar{\chi}(x) U_\nu(x) T^a \chi(x+\hat{\nu}a) \right], 
	\label{eq:eps-current} \\
	P^a_P(x) & = & \varepsilon(x) \bar{\chi}(x) T^a \chi(x) ,
	\label{eq:eps-pseudoscalar}
\end{eqnarray}
satisfying PCAC relations~\cite{Smit:1987zh} (for equal renormalized 
masses~$m$)%
\footnote{In the flavor singlet $\tr T^a\neq0$, but the last term
vanishes in a average over a hypercube, so it is unimportant.}
\begin{equation}
	a^{-1} \sum_\nu [A^{\nu a}_P(x)-A^{\nu a}_P(x-\hat{\nu}a)] =
		2mP^a_P(x)+2a^{-4}\varepsilon(x) \tr T^a.
	\label{eq:PCAC}
\end{equation}
It is tempting to say that the axial anomaly does not appear in
Eq.~(\ref{eq:PCAC}) owing to Eq.~(\ref{eq:anomaly}), with each
contribution at quarter strength.
It is better, however, to note that it is not expected,
because the symmetry (\ref{eq:axial}) is exact.
The axial anomaly is also undesired.
Under shifts $A^{\nu a}_P$ and $P^a_P$ transform~as
\begin{equation}
	S_\mu: \left\{ %}
	\begin{array}{rcl}
		A^{\nu a}_P(x) & \mapsto & 
			(-1)^{n_\mu} A^{\nu a}_P(x+\hat{\mu}a) \\
		P^a_P(x) & \mapsto & (-1)^{n_\mu} P^a_P(x+\hat{\mu}a) 
	\end{array} \right.,\quad \forall \mu,\nu,
	\label{eq:shiftP}
\end{equation}
so one sees that these bilinears are \emph{not} taste singlets, rather 
they transform under the (non-singlet) $P$ irrep.
This non-singlet character of $A^{\nu a}_P$ and $P_P$ provides a clue that 
the new quantum number taste plays a key role in the physics that emerges 
in the continuum limit.

The change of variables~(\ref{eq:KawamotoSmit}) followed by the projection 
from $\Chi^i_\alpha$ to $\chi^i$ reduces the number of species from 16 
to~4.
The action can be rewritten using physically suggestive fields
\begin{equation}
	\psi_{\alpha t}(y) = \case{1}{8} \sum_r \Omega_{\alpha t}(r) 
		U(y,y+r) \chi(y+r), \quad
	\bar{\psi}_{\alpha t}(y) = \case{1}{8} \sum_r \bar{\chi}(y+r)
		\Omega^\dagger_{t \alpha}(r) U(y+r,y) ,
\end{equation}
where $y$ labels hypercubes of size $2^4$ or, equivalently, sites on a 
coarser lattice of spacing~$b=2a$, $r$~runs over the hypercube, and 
$U(y,y+r)$ is parallel transport along some chosen path from $y$ to $y+r$.
As in Eq.~(\ref{eq:mixed-bilinear}), the indices $\alpha$ and $t$ are 
interpreted as Dirac and taste indices.
Suppressing the gauge field, the action is then 
rewritten~\cite{Gliozzi:1982ib,KlubergStern:1983dg} 
\begin{eqnarray}
	S_{\mathrm{stag}} & = &
		\case{1}{2} b^3 \sum_{y,\mu} \sum_{t=1}^4 \bar{\psi}_t(y) \gamma_\mu
		\left[\psi_t(y+\hat{\mu}b) - \psi_t(y-\hat{\mu}b)\right] +
		m_0 b^4 \sum_y \sum_{t=1}^4 \bar{\psi}_t(y) \psi_t(y) \nonumber \\
		& - & \case{1}{2} b^3 \sum_{y,\mu} \sum_{t,t'=1}^4 \bar{\psi}_t(y) 
			\gamma_5 [\xi_5 \xi_\mu]_{tt'} \left[ \psi_{t'}(y+\hat{\mu}b) + 
			\psi_{t'}(y-\hat{\mu}b) - 2\psi_{t'}(y) \right],
	\label{eq:taste-basis}
\end{eqnarray}
which looks like a lattice field theory of four Dirac fermions,
with a Wilson-like term to alleviate doublers.
For perturbative gauge fields this appearance continues to hold. 
For example, the beta-function starts with 
$\beta_0=\case{11}{3}N_c-\case{2}{3}4n_f$ \cite{Sharatchandra:1981si},
and the exact $\Gamma_4$ part of the shift symmetry ensures that mass 
renormalization is taste-independent~\cite{Golterman:1984cy,Mason:2005bj}.

The central conjecture of (unrooted) staggered fermions is that the
picture of four (or $4n_f$) Dirac fermions holds non-perturbatively.
I say ``conjecture'' to mathematicians because it is not proven
rigorously (and I could say the same about all formulations of lattice
fermions), and to physicists because the coupling to gauge fields is
based on the one-component action, Eq.~(\ref{eq:staggered}).
Consequently the spacetime and flavor-like symmetries are entangled, so
it is not especially transparent how SO(4) rotational symmetry and
${\rm SU_V}(4n_f)\times{\rm SU_A}(4n_f)\times{\rm U_V}(1)$ 
flavor-taste chiral symmetry emerge in the continuum limit. 

To get an idea of the complications (see, \emph{e.g.},
Refs.~\cite{Golterman:1984cy,Kilcup:1986dg} for details), let us 
consider some important cases.
The hypercubic rotations (denoted ${\rm SW}_4$) are embedded
\begin{equation}
	{\rm SW}_4  \subset [{\rm SO}(4)\times {\rm SO}(4)]_{\rm diag} \subset
		{\rm SO}(4) \times {\rm SU_V}(4) \subset
		{\rm SO}(4) \times {\rm SU_V}(4n_f)
	\label{eq:embedrotation}
\end{equation}
into the diagonal subgroup of the Euclidean group and a flavor-singlet 
subgroup of the ${\rm SU_V}(4n_f)$ group.
The Clifford group abstracted from the shifts and the vector flavor
symmetries are embedded
\begin{equation}
	\Gamma_4 \times {\rm SU_V}(n_f) \subset {\rm SU_V}(4n_f)
	\label{eq:embedtaste}
\end{equation}
in the vector flavor-taste symmetries.
The axial flavor symmetries are embedded
\begin{equation}
	{\rm U_A}(n_f) \subset {\rm SU_A}(4n_f)
	\label{eq:embedaxial}
\end{equation}
in the flavor-taste non-singlet axial symmetries.
For sect.~\ref{sec:anti_Creutz}, the embedding~(\ref{eq:embedaxial})
along with Eqs.~(\ref{eq:eps-current})--(\ref{eq:shiftP}) are key.
The exact chiral symmetries are \emph{not} taste singlets.
Therefore, they cannot have anything to do with anomalies, gauge-field
topology, zero modes of the Dirac operator, or their consequences, such
as 't~Hooft vertices.

If the emerging picture is correct, staggered fermions must possess a
bilinear corresponding in the continuum limit to the flavor- and
taste-singlet axial current.
Analogously to the case of Wilson fermions, this is not a Noether
current, but the explicit breaking is superficially of order~$a$.
The anomalous current extends over a whole 
hypercube~\cite{Sharatchandra:1981si,Smit:1987zh}:
\begin{eqnarray}
	A^\mu_I(x) = \frac{i}{32}\sum_{b+c=d} 
		\eta^\mu(x+c) & & \hspace{-1.5em} \nonumber
		\eta_1(x+b)\eta_2(x+b)\eta_3(x+b)\eta_4(x+b)
		\left[ \vphantom{U_\mu^\dagger} \right. \\[-0.7em]
		&   & \bar{\chi}(x+c+\hat{\mu}a)
		U^\dagger_\mu(x+c)\bar{U}(x+c,x+b)\chi(x+b) \nonumber \\
		& - & \left. \bar{\chi}(x+b)\bar{U}(x+b,x+c)
		U_\mu(x+c)\chi(x+c+\hat{\mu}a) \right], 
	\label{eq:current}
\end{eqnarray}
where $d=(\hat{1}+\hat{2}+\hat{3}+\hat{4})a$, $\bar{U}(x+b,x+c)$ is the 
average of parallel transport over paths from $x+b$ to $x+c$, and the sum 
is over all 4-dimensional diagonals of the hypercube at $x$.
Under shift symmetry $A^\mu_I$ is a taste singlet, as is the corresponding 
pseudoscalar density
\begin{equation}
	P_I(x) = \frac{i}{16}\sum_{b+c=d} 
		\eta_1(x+b)\eta_2(x+b)\eta_3(x+b)\eta_4(x+b)
		\bar{\chi}(x+b)\bar{U}(x+b,x+c) \chi(x+c).
	\label{eq:pseudoscalar}
\end{equation}
$A^\mu_I$ and $P_I$ satisfy a PCAC relation~\cite{Smit:1987zh} with the
anomaly of four Dirac fermions~\cite{Sharatchandra:1981si}.
Compared to Eq.~(\ref{eq:PCAC}) a complication is the need for 
renormalization~\cite{Smit:1987zh}, but this is analogous to the 
renormalization of Ward-Takahashi identities of Wilson fermions.

\begin{figure}[b]
	\centering
	\includegraphics[width=\textwidth]{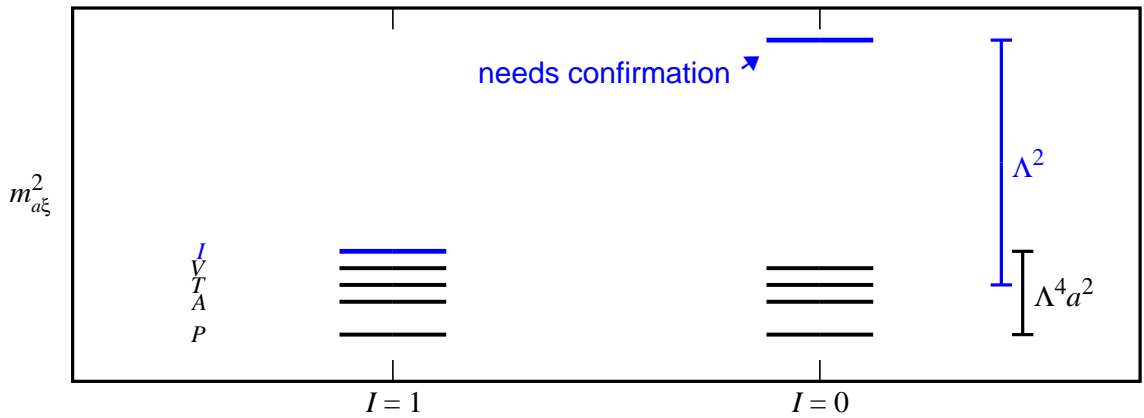}
	\caption[fig:pseudoscalar]{Pseudoscalar spectrum for 2 flavors of
	staggered fermion, so 8 species in all.
	The isovector multiplets (with $I_3=+1$, $0$, $-1$) each
	consist of sixteen states, split by lattice artifacts of order
	$\Lambda^4a^2$ into submultiplets with 1, 4, 6, 4, and 1 states 
	(for, respectively, irreps of taste $I$, $V$, $T$, $A$, and $P$).
	The isosinglet multiplet is similar, except that the
	taste-singlet~$I$ splits from the others by continuum-QCD effects
	(as usual).
	This $\eta'$-like state suffers from noisy correlators, and
	numerical data are consistent with this picture without being
	definitive~\cite{Gregory:2007ev}.}
	\label{fig:pseudoscalar}
\end{figure}

If staggered fermions do indeed provide a ($4n_f$)-species version of QCD, 
then general features of the hadron spectrum should provide clear 
numerical evidence.
The chiral symmetries, and the standard line of reasoning for 
spontaneously broken symmetry, imply that the pseudoscalar meson masses 
should behave (with degenerate quark masses $m_q$) as
\begin{equation}
	m^2_{a\xi} = 2m_qB + a^2\Delta_\xi + \mu^2\delta_{a0}\delta_{\xi I},
	\label{eq:mass2}
\end{equation}
where $a=0,\ldots,n_f^2-1$ labels flavor and $\xi\in\{P, A, T, V, I\}$ 
labels taste.
The dynamical quantities $B$, $\Delta_\xi^{1/4}$, and $\mu$ are of 
order~$\Lambda$, the characteristic dynamical scale of the gauge fields.
The exact (taste-nonsinglet) chiral symmetries 
of Eq.~(\ref{eq:axial})
imply $\Delta_P=0$.
The ``$\eta'$-$\pi^0$ splitting'' parameter~$\mu^2$ can be generated only 
for the flavor- \emph{and} taste-singlet meson.
This pattern is shown (for $n_f=2$) in Fig.~\ref{fig:pseudoscalar}.
It is consistent with extensive numerical simulations in the quenched
approximation, as well as with 2 and 2+1 flavors of (rooted, staggered)
sea quarks~\cite{Bernard:2001av,Gregory:2007ev}, including the scaling 
of the pseudoscalar splittings $a^2\Delta_\xi$~\cite{Bernard:2006wx}.

The approach to the continuum limit can be clarified with 
two theoretical tools, the
Symanzik effective field 
theory~\cite{Lepage:1998vj,Lee:1999zxa} and chiral perturbation theory 
($\chi$PT)~\cite{Lee:1999zxa,Bernard:2001yj,Aubin:2003mg,Sharpe:2004is}.
In the Symanzik local effective Lagrangian (LE$\mathcal{L}$), 
dimension-four interactions have SO(4) rotational symmetry and, by design,
${\rm SU_V}(4n_f)\times{\rm SU_A}(4n_f)$ flavor-taste chiral symmetry, 
softly broken by mass terms.
Dimension-six four-quark operators break these down to ${\rm SW}_4$ and
$\Gamma_4\times{\rm SU_V}(n_f)\times{\rm U_A}(n_f)$, as indicated by the 
embeddings~(\ref{eq:embedrotation})--(\ref{eq:embedaxial})
\cite{Lee:1999zxa,Bernard:2001yj,Aubin:2003mg}.
In particular, some of these operators are invariant only under the 
$\Gamma_4$ taste symmetry, not a full SU(4).
They are variously called taste-breaking or taste-exchange interactions.
The LE$\mathcal{L}$ analysis clarifies why smearing strategies reduce
the strength of taste-symmetry breaking~\cite{Blum:1996uf}.
If successful, smeared actions should yield smaller splittings $\Delta_\xi$.
They do~\cite{Orginos:1999cr}.
They also improve the scaling of the $\rho$-meson mass and the static
potential~\cite{Bernard:1999xx}.

The splittings and other effects of broken taste symmetry are, of
course, a~complication.
They lead to multi-parameter fitting procedures, for example of the 
pseudoscalar decay constants~\cite{Aubin:2003uc}.
Another example is the $\pi$-$\pi$ threshold, which splits into five 
$\pi_\xi$-$\pi_\xi$ thresholds, $\xi\in\{P,A,T,V,I\}$.
In last year's plenary talk on this subject~\cite{Sharpe:2006re}, the
need for complicating fitting was one reason to deem lattice QCD with
rooted staggered quarks ``ugly.''
Many of the complications come not from rooting but from the intricate
symmetry structure and the desire for statistically sound fits.
Fitting is nicely illustrated by $f_\pi$ and $f_K$.
The staggered data start so close to the experimental result that just
about any chiral extrapolation would agree with experiment, but
statistically good fits are obtained only when staggered $\chi$PT is
used~\cite{Aubin:2004fs}.
Thus, these fits provide further evidence that staggered (valence)
fermions do indeed simulate $4n_f$ flavors of quark.

To study how these properties of staggered fermions relate to gauge-field 
topology, one must consider the eigenvalues of the staggered Dirac 
operator.
The exact (non-singlet) chiral symmetries imply that eigenvalues come in
complex conjugate pairs, $\pm i\lambda_i+m_0$, with orthonormal
eigenvectors $f_i(x)$ and $\varepsilon(x)f_i(x)$.
In a gauge field with a non-zero topological charge $Q$, the continuum
Dirac operator has $n_\pm$ zero modes of chirality $\pm1$, satisfying
the index theorem $n_+-n_-=n_sQ$, where $n_s$ is the total number of
species.
With lattice staggered fermions, the zero modes are no longer exact.
Nevertheless, if staggered fermions do indeed yield four quarks, then 
several features must emerge dynamically~\cite{Smit:1986fn}.
First, low-lying eigenvalues must cluster in quartets.
Second, some (pairs of) eigenvalues should be near-zero modes with 
$\lambda_i\sim\Lambda^2a$.
Third, the chirality of the corresponding eigenvectors should be close to 
$\pm1$, in a way that satisfies the index theorem.
Finally, in the finite-volume $\epsilon$ regime, the distributions of
eigenvalues should agree with expectations from random matrix
theory~\cite{Damgaard:2001ep}.

Early numerical work was inconclusive~\cite{Vink:1988ss}, but with
improved staggered actions, quartets of eigenvalues clearly emerge in
numerical simulations~\cite{Follana:2004sz}.
In particular, if the gauge field has topological charge $Q$, then 
the eigenvalue spectrum contains quartets of the same (taste-singlet, 
Eq.~(\ref{eq:pseudoscalar})) chirality, such that the index theorem is 
satisfied~\cite{Follana:2004sz}.
Even with smearing, early comparisons of random matrix theory suggested
that staggered fermions are topology blind~\cite{Damgaard:1999bq}.
Equation~(\ref{eq:mass2}) implies two regimes of
interest~\cite{Damgaard:2001fg,Wong:2004nk}, however,
\begin{eqnarray}
	\epsilon~\textrm{regime}:\hspace{1.8em}
		\Lambda & \gg L^{-1} \gg & m_\xi,\; \forall \xi,
	\label{eq:epsilon}  \\
	\epsilon'~\textrm{regime}:\; m_{\xi\neq P} & \gg L^{-1} \gg & m_P.
	\label{eq:epsilon'}
\end{eqnarray}
In the $\epsilon$ regime all ``pions'' are pseudo-Goldstone bosons, 
whereas in the $\epsilon'$ regime most of them are very massive particles.
In the $\epsilon$ regime staggered fermions should and do behave like a 
4-species theory~\cite{Wong:2004nk}; in the $\epsilon'$ regime they
should not~\cite{Damgaard:2001fg} and do~not~\cite{Damgaard:1999bq}.

In summary, na\"ive fermions appear to be problematic, because the
anomaly is not generated.
Analysis of the unexpected doubling symmetries rescues the formulation,
however, yielding staggered fermions.
Spacetime and flavor-like symmetries become entangled, making the
interpretation as Dirac fermions less transparent.
But now not only are the exact chiral symmetries non-singlets under a
flavor-like quantum number called taste, but also an anomalous
taste-singlet axial current can be found.
On this basis, the emergence of $4n_f$ Dirac fermions is theoretically 
plausible.
Owing to a wealth of results examining the nonperturbative content and 
structure, it is also fair to say that the validity of (unrooted) 
staggered fermions has been established numerically.

\section{Rooting with full SU(4) taste symmetry}
\label{sec:rooting}

The previous section reviewed some aspects of staggered fermions,
focusing on how four species emerge in the continuum limit.
Before discussing whether the fourth-root procedure can reduce these
four species to one, I would like to present a \emph{Gedanken} algorithm.
The aim is to separate some potentially confusing aspects of rooting,
free of the complications of staggered fermions' taste-exchange
interactions.

Suppose that an algorithm designer with a good imagination (and a wicked
sense of humor) found a way to speed up ``your favorite lattice
fermions'' by substituting
\begin{equation}
	{\det}_1(\Dirac + m) = \left\{
		{\det}_4[(\Dirac + m) \otimes \bm{1}_4] \right\}^{1/4},
	\label{eq:cleverBoy}
\end{equation}
thereby introducing four ``tastes.''
Here $\det_1$ is a determinant for $n_f$ flavors, with 1~taste per
flavor; $\det_4$ is for $n_f$ flavors, but 4~tastes per flavor.
If the determinant is real and positive, this step does not 
change the Monte Carlo weight at all, because the right-hand side is 
just a clever trick to calculate the left-hand side.
Of course, the trick fails if 
the left-hand side is not real
(\emph{e.g.}, for nonzero chemical potential~\cite{Golterman:2006rw}), 
or if it can be negative (\emph{e.g.}, $m<0$).

To probe the dynamics of this system, first introduce sources
$(J^a,J^a_5)$ for scalar and pseudoscalar meson operators
$(-i\bar{\psi}T^a\psi,-i\bar{\psi}T^a\gamma_5\psi)$ in
the original theory:
\begin{eqnarray}
	Z[J^a,J^a_5] & = & \int\mathcal{DU}\, \{
		{\det}_1(\Dirac + m + J + J_5\gamma_5)\}^{N_r} 
	\label{eq:originalZ} \\
		& = & \int\mathcal{DU}\, \left\{
		{\det}_4[(\Dirac + m + J + J_5\gamma_5) \otimes \bm{1}_4] 
		\right\}^{N_r/4} ,
	\label{eq:cleverZ}
\end{eqnarray}
where $\mathcal{DU}$ is the gauge-field measure (including
$e^{-S_{\mathrm{gauge}}}$), 
and the number of ``replicas'' $N_r$ will be useful below.
The $T^a$ are flavor matrices (with $T^0=i\bm{1}/\sqrt{2n_f}$), 
and $J_{(5)}=-iJ^a_{(5)}T^a$.

Spontaneous symmetry breaking is revealed by looking at the Legendre 
effective action
\begin{equation}
	e^{\Gamma(\sigma^a,\pi^a)} = \left. Z[J,J_5] 
		e^{-\sum_x[J^a(x)\sigma^a(x)+J^a_5(x)\pi^a(x)]} 
		\right|_{J_{(5)}=J_{(5)}(\sigma,\pi)},
	\label{eq:Legendre}
\end{equation}
where $J_{(5)}(\sigma,\pi)$ are defined implicitly, so that they 
create prescribed fields $(\sigma^a,\pi^a)$:
\begin{equation}
	\sigma^a(x) = \left.\frac{1}{Z} \frac{\partial Z}{\partial J^a(x)}
		\right|_{J_{(5)}=J_{(5)}(\sigma,\pi)}, \quad
	\pi^a(x) = \left.\frac{1}{Z} \frac{\partial Z}{\partial J_5^a(x)}
		\right|_{J_{(5)}=J_{(5)}(\sigma,\pi)}.
	\label{eq:source-for-field}
\end{equation}
The mass matrices for $\sigma^a$ and $\pi^a$ are obtained from second
derivatives of $\Gamma$.
In the case at hand, they are constrained by symmetry---invariance of 
the fermion action is expressed in Eqs.~(\ref{eq:originalZ}) 
and~(\ref{eq:cleverZ}) as the invariance of the determinants under a 
change of basis.
One finds~\cite{Chodos:2000dn}
\begin{eqnarray}
	\left.
	\frac{\partial^2\Gamma}{\partial\sigma^a\partial\sigma^c} f^{cdb} \sigma^d
	\right|_{\sigma_0,\pi_0} +
	\left.
	\frac{\partial^2\Gamma}{\partial\sigma^a\partial\pi^c}    f^{cdb} \pi^d
	\right|_{\sigma_0,\pi_0} & = & 0,
	\label{eq:sigmaMass} \\
	\left.
	\frac{\partial^2\Gamma}{\partial\pi^a\partial\pi^c} d^{cdb} \sigma^d
	\right|_{\sigma_0,\pi_0} +
	\left.
	\frac{\partial^2\Gamma}{\partial\pi^a\partial\sigma^c} d^{cdb} \pi^d
	\right|_{\sigma_0,\pi_0} & \propto & \textrm{mass and anomaly terms},
	\label{eq:pionMass}
\end{eqnarray}
where the vacuum fields $(\sigma_0,\pi_0)$ are those in the absence of 
sources $J_{(5)}$.\footnote{%
Instead of fixing $J_{(5)}$ in Eq.~(\ref{eq:source-for-field}) to get
prescribed fields, simply set $J_{(5)}=0$ on the right-hand side to
get $(\sigma_0,\pi_0)$.}
If one assumes that all vacuum fields vanish except the flavor-singlet 
scalar $\sigma_0^0$, then one obtains a constraint
\begin{equation}
	\left. \frac{\partial^2\Gamma}{\partial\pi^a\partial\pi^b}
	\right|_{\sigma_0,\pi_0} \propto \textrm{mass and anomaly terms},
	\label{eq:pion}
\end{equation}
and no constraint on $\partial^2\Gamma/\partial\sigma^a\partial\sigma^b$.
Equation~(\ref{eq:pion}) is the basis for formulae such as
Eq.~(\ref{eq:mass2}).
Of course, the dynamical assumption that $\sigma_0^0\neq0$ has not been 
proven mathematically but has been established numerically with various
types of lattice fermions.

Equation~(\ref{eq:cleverZ}) apparently has more symmetry than
Eq.~(\ref{eq:originalZ})---${\rm SU}(4n_f)\times{\rm SU}(4n_f)$ instead
of ${\rm SU}(n_f)\times{\rm SU}(n_f)$.
I will call the extended symmetry a \emph{phantom} symmetry, because it
is a figment of an algorithm designer's imagination.
One can study the dynamical consequences of the phantom symmetry by
promoting the sources to taste-nonsinglets~\cite{Bernard:2006vv}:
\begin{equation}
	Z[J^A,J^A_5] = \int\mathcal{DU}\, \left\{
		{\det}_4[(\Dirac + m) \otimes \bm{1}_4 + J + J_5\gamma_5] 
		\right\}^{N_r/4} ,
	\label{eq:clever5}
\end{equation}
where the flavor-taste generator index~$A$ now runs from 0 to $(4n_f)^2-1$.
One finds the same pattern of spontaneous symmetry breaking as in
Eqs.~(\ref{eq:sigmaMass})--(\ref{eq:pionMass}), leading to mass
relations like (\ref{eq:pion}) also for the taste nonsinglet phantom
pseudoscalars.
In all, there are $(4n_f)^2-1$ pseudo-Goldstone bosons, instead of just 
$n_f^2-1$, and one taste- and flavor-singlet $\eta'$-like meson.

But what should one make of the phantoms?
The equivalence of the two formulations, (\ref{eq:originalZ}) and 
(\ref{eq:cleverZ}), shows that phantoms do not exist.
The symmetry analysis suggests they do.
(A~numerical simulation of your favorite rooted fermions could look at
taste-nonsinglet correlators to decide and would probably find them.)
The conundrum is easily resolved if the phantoms violate unitarity and
cancel each other out in correlators that are oblivious to
taste~\cite{Bernard:2006vv}.
The lack of unitarity, particularly when constrained by symmetry, 
is not a concern and, in fact, is familiar in gauge 
theories~\cite{Itzykson:1980rh}.

The \emph{Gedanken} algorithm helps establish a foundation for some 
questions concerning rooted staggered fermions.
First, we see what kinds of correlators are physical.
For bosonic taste, these are taste singlets and anything related to
taste singlets by symmetry.%
\footnote{For example, one may use the taste-$P$ irrep for pseudoscalar
mesons, the taste-$V$ irrep for vector mesons, \emph{etc}.}
For fermionic taste, one needs single-taste
correlators~\cite{Sharpe:2006re}.
It is possible to construct \emph{unphysical} correlators, which could
lead to confusion or mistakes.
Another use of Eqs.~(\ref{eq:originalZ}) and (\ref{eq:cleverZ}) is to
set a criterion for proving staggered fermions \emph{incorrect}.
Any argument that would also kill the \emph{Gedanken} algorithm simply 
does not pertain to the issues at hand.
An example is the proof that $(\Dirac+m)^{1/4}$ is
non-local~\cite{Bunk:2004br}, which does not prove that
$[\det_4(\Dirac+m)]^{1/4}$ cannot be obtained from a path integral with
a local action~\cite{Adams:2004mf}.
Attempts to disprove staggered fermions must focus on features not
shared with the \emph{Gedanken} algorithm, namely the violations of
perfect SU(4) taste symmetry.
Finally, note that the details of the phantom sector depend on $N_r$.
For $N_r\in\mathbb{Z}$, the rooted formulation of 
Eq.~(\ref{eq:cleverZ}) is identical to a manifestly sound quantum 
field theory.
On the other hand, Eqs.~(\ref{eq:originalZ}) and (\ref{eq:cleverZ}) show
that irrational choices, like $N_r=\pi$, are simply irrational.

\section{Rooting with staggered fermions}
\label{sec:synthesis}

We are now ready to discuss rooted staggered fermions, which is the
\emph{Ansatz} that quarks can be simulated with Eq.~(\ref{eq:rooting}),
using any of several available
algorithms~\cite{Fucito:1980fh,Clark:2006wp}.
An initial set of arguments in favor are as follows.
First, Eq.~(\ref{eq:rooting}) resembles Eq.~(\ref{eq:clever5}),
especially when one looks at the $\Dirac_{\mathrm{stag}}$ defined via
Eq.~(\ref{eq:taste-basis}); the four-taste determinant has SU(4)
taste-violating parts that are suppressed superficially by powers of the
lattice spacing.
Second, perturbative renormalization~\cite{Golterman:1984cy,Mason:2005bj} 
and the nonperturbative features discussed at the end of
sect.~\ref{sec:staggered} support the picture of unrooted staggered
fermions as a QCD-like theory, in which violations of SU(4)
taste symmetry vanish in the continuum limit.
Finally, having thus achieved a $4\times4$ block structure, the fourth
root is as straightforward as in sect.~\ref{sec:rooting}.

Implicit in this line of reasoning is the assumption that the continuum 
limit and rooting commute.
We shall come back to this below.
First, however, I would like to consider objections that hold even if 
these arguments are all correct.
In particular, I stipulate for now that SU(4$n_f$) flavor-taste symmetry
emerges in the continuum limit, citing the evidence presented at the end
of sect.~\ref{sec:staggered}.

In the context of rooting most of the SU(4$n_f$) symmetry is a phantom
symmetry.
As such one should ask whether a phantom sector of particles is seen in 
numerical simulation. \nopagebreak
(They are, in the scalar propagator, see below.)
If so, they violate unitarity. \pagebreak
Moreover, at non-zero lattice spacing the symmetry is smaller than
SU($4n_f$), as exhibited in (\ref{eq:embedtaste}) and
(\ref{eq:embedaxial}).
Consequently, the phantoms are not degenerate, so they will not cancel
identically in physical correlators.
Fortunately, sect.~\ref{sec:rooting} shows in general (if sketchy) terms
that the rooted dynamics provide a safe house for phantoms, when (as
stipulated above) they become degenerate in the continuum limit.
Then the phantom sector is benign, even if it is not unitary, because
the cancellation becomes perfect.

The ramifications of non-degenerate phantoms are still ugly, in some 
cases perhaps even scary.
A useful and illuminating tool is $\chi$PT adapted for the case at hand:
rooted, staggered chiral perturbation theory
(RS$\chi$PT)~\cite{Bernard:2006zw}.
This extension of partially quenched $\chi$PT encodes the $N_r$
dependence, so it is sensitive to the dynamics of the sea.
As mentioned in sect.~\ref{sec:staggered}, the fit to pseudoscalar meson
masses and decay constants provides clear evidence that this description
works.
To isolate the sensitivity of the numerical data to the sea quarks, 
MILC now tries fits in which the number of replicas floats.
At this conference Bernard reported~\cite{Bernard:2007ps}
(in his notation, $N_r=4n_r$)
\begin{equation}
	\case{1}{4}N_r = n_r = 0.28(2)(3),
	\label{eq:nrFit}
\end{equation}
in striking agreement with the expectation $1/4$.
Further evidence comes from heavy-light decay constants.
Continuum partially quenched $\chi$PT blatantly fails to describe the
numerical data~\cite{Simone:2004xpt}, while RS$\chi$PT yields a
statistically sound fit~\cite{Aubin:2005ar}.
The topological susceptibility is also better-described with RS$\chi$PT
than with continuum $\chi$PT~\cite{Billeter:2004wx}.

The pseudoscalar mesons just mentioned are low-lying states, but
higher mass particles must be considered too.
Correlators, for example those of glueballs, contain not only the single
hadron of interest but also two-phantom-particle states.
It can become hard to determine the single-particle mass, but these
correlators are still interesting, because the two-particle contribution
probes the sea~\cite{Prelovsek:2005rf}.
Correlators of taste-singlet scalar mesons contain $\pi_\xi$-$\eta_\xi$
pairs, $\xi=P,A,T,V$, and $\pi_\xi$ and $\eta_\xi$ consist of a valence
and a sea quark.
These $\pi_\xi$-$\eta_\xi$ bubbles can be described with $\chi$PT,
analogously to the quenched scalar correlator~\cite{Bardeen:2001jm}.
The bubbles' weights depend on $N_r$.
For $N_r<4$ some weights are negative, as one expects when unitarity is
violated.
Once all thresholds become degenerate, the bubbles always add up to
the correct total.
Comparisons of numerical data with the RS$\chi$PT formulae confirm this
picture~\cite{Bernard:2007qf}, supporting the correctness of the rooted
sea.
A similar story holds in the case of a mixed action with a rooted sea 
and Ginsparg-Wilson valence quarks; the numerical data are again well 
described by mixed-action $\chi$PT~\cite{Aubin:2007wr}.

In summary, the violations of unitarity are a serious objection and can 
be a practical obstacle.
The numerical data suggest, however, that these effects, including 
their influence on the error budget, can be managed with RS$\chi$PT.
Furthermore, the analysis of sect.~\ref{sec:rooting} reveals a safe 
house for phantoms, not only for pseudoscalar mesons, but also for 
other hadrons.

A distinct, but related, issue is the locality of rooted staggered 
fermions.
Critics of the rooting procedure have long suspected non-local
behavior~\cite{Jansen:2003nt}, but to my knowledge the concern is that a
Lagrangian of the form $\bar{\psi}(\Dirac+m)^{1/4}\psi$ is
non-local~\cite{Bunk:2004br}.
Because this line of attack would kill the manifestly correct
\emph{Gedanken} algorithm, this concern is a red herring.
There is, however, another kind of non-local behavior.
Suppose one writes
\begin{equation}
	{\det}_4(\Dirac_{\rm stag}+m) = 
		\left[{\det}_1(\Dirac_{\rm SU(4)}+m)\right]^4 T,
	\label{eq:factors}
\end{equation}
\nopagebreak
where the first factor on the right-hand side is manifestly SU(4) 
symmetric.
The remaining factor~$T$ cannot represent a set of local interactions \pagebreak
for the gauge field, because the left-hand side generates a spectrum
with taste splittings, and the SU(4)-symmetric determinant does
not~\cite{Bernard:2006ee}.
It is certainly accurate to call this a non-locality, but I am not sure 
how illuminating it is.
Before taking the fourth root, the non-locality is particle-like.
After taking the fourth root on both sides of Eq.~(\ref{eq:factors}), 
it strikes me as plausible that the non-locality of $T^{1/4}$ does not 
ruin cluster decomposition, it may become local in the continuum 
limit~\cite{Adams:2003rm}, and it certainly is not the non-locality of
a propagator $(\Dirac+m)^{-1/4}$.
Thus, it is not the non-locality that critics seem to have in mind.

To be confident that the continuum limit is correct, it would be helpful
if one could establish a Symanzik LE$\mathcal{L}$, not with an Ansatz
based on symmetries, but through a derivation from the underlying
lattice field theory.
The Symanzik structure is evident at the tree level, and it seems to be
accepted without controversy at all orders in perturbation theory.
(It is not, however, proven at the level needed to prove Reisz's
theorems~\cite{Reisz:1987pw}.
An important first step has been to extend the power-counting theorem to
staggered fermions~\cite{Giedt:2006ib}.)
In a forthcoming paper, Bernard, Golterman, and Shamir (BGS) show a new
way to do~so.
For a short version with more details than I can give here, see
Golterman's talk at this conference~\cite{Bernard:2007yt}.

The new work of BGS is based on Shamir's block-spin renormalization 
group~\cite{Shamir:2004zc}, which was covered thoroughly in Sharpe's 
review~\cite{Sharpe:2006re} last year.
After $n$ blocking steps, Shamir arrives at a blocked staggered operator
\begin{equation}
	(\Dirac+m)_n \otimes \mathbf{1}_4 + a\Delta_n
\end{equation}
with a SU(4)-symmetric part $(\Dirac+m)_n\otimes\mathbf{1}_4$ and a 
taste-breaking defect $\Delta_n$.
The blocked quantities live on a lattice with spacing $2^na$, and
in the continuum limit $n\to\infty$, $a\to0$, $a_c=2^na$ fixed.
A~power-law divergence in $\Delta_n$ is expected to be $a_c^{-2}$, 
implying that the combination $a\Delta_n$ vanishes in the continuum 
limit.

The blocked determinant resembles Eq.~(\ref{eq:clever5}), with sources
set to provide $a\Delta_n$, and $\mathcal{DU}$ the complicated measure
for the gauge field in Ref.~\cite{Shamir:2004zc}.
BGS develop an expansion in the underlying spacing~$a$ with the aid of
some bookkeeping devices.
The determinant is generalized to be
\begin{equation}
	\{{\det}_4[(\Dirac+m)_n \otimes \mathbf{1}_4 + a\Delta_n]\}^{n_r} \to
	\{{\det}_1[(\Dirac+m)_n]\}^{N_s} \frac{% 
	\{{\det}_4[(\Dirac+m)_n \otimes \mathbf{1}_4 +ta\Delta_n]\}^{n_r}}{%
	\{{\det}_4[(\Dirac+m)_n \otimes \mathbf{1}_4]\}^{n_r}},
	\label{eq:BGS}
\end{equation}
where $N_s$ need not equal $N_r=4n_r$, and $t$ need not equal~$1$.
Sources for valence fermions are
\begin{equation}
	\exp\left\{\bar{\eta}\left[(\Dirac+m)_n \otimes \mathbf{1}_4
		 +va\Delta_n \right]^{-1} \eta % \otimes \mathbf{1}_{n_r} \eta 
		 \right\},
	\label{eq:source}
\end{equation}
where, again, for bookkeeping $v$ need not equal~$1$.
Post-analysis, one may set $N_s=N_r$, $t=v=1$.

Expressions~(\ref{eq:BGS}) and~(\ref{eq:source}) can be expanded in $t$
and $v$, justified by the small quantity $a\Delta_n$.
The double-expansion can then be reverse-engineered to reproduce the
Lee-Sharpe LE$\mathcal{L}$ for staggered fermions~\cite{Lee:1999zxa}.
Each bookkeeping parameter brings an advantage.
Possible nonperturbative nonpolynomial dependence of the LE$\mathcal{L}$ 
on $N_s$ is kept, because other factors are expanded in $t$ and~$v$.
The expansion in $t$ ensures that, to any order in $a$, the dependence
on $n_r$ is polynomial, so analytical continuation from an integer to
$1/4$ is allowed.
The expansion in $v$ shows that valence quarks control the symmetries
and the field content of the LE$\mathcal{L}$, which do not depend
on~$n_r$.
Pending confirmation of the assumptions built into the 
blocking~\cite{Shamir:2004zc,Sharpe:2006re}, these are a strong results.

This analysis underscores the importance of checking numerically that 
$\Delta_n$ scales in such a way as to justify the expansions.
There are several pieces of evidence to suggest it does.
Exhibit~A is all experience with anomalous dimensions in QCD. \pagebreak
Exhibit~B is all experience with the pseudoscalar spectrum as taste-breaking 
of valence fermions is reduced~\cite{Orginos:1999cr,Bernard:2001av};
heuristically this is like reducing $v$ without reducing~$a$.
Lastly, Exhibit~C is a pilot investigation of~$\Delta_n$
itself~\cite{Bernard:2005gf}.

A lack of scaling of $a\Delta_n$ would kill rooted staggered fermions,
because then we would be left with a non-unitary, non-local theory, even
in the continuum limit.
It is therefore worth stating how the evidence given above could be 
misleading.
The pilot investigation~\cite{Bernard:2005gf} is not yet definitive.
The pseudoscalar splittings probe only four of the plethora of 
taste-exchange effects generated by $\Delta_n$.
Finally, it could be that rooting and the continuum limit do not 
commute, in a way so profound that the rooted determinant itself
generates $\Delta_n$'s anomalously large anomalous
dimension~\cite{Sharpe:2006re}.
Such ``self-inconsistency''  is possible, but highly implausible.

\section{Explicit refutation of Refs.~\Creutz}
\label{sec:anti_Creutz}

With the preceding sections' outline of rooted staggered quarks as a
basis for discussion, I can now address Mike Creutz's specific
criticisms, which below are summarized in \textsl{slanted font}.
Many of my counterarguments are the same as in the original 
refutation~\cite{Bernard:2006vv}.

\subsection{Order of limits}
\label{sub:limits}
\noindent\textsl{Rooted staggered fermions require a ``peculiar'' order
of limits, $a\to0$ with $m$~fixed, followed by
$m\to0$~\cite{Creutz:2006ys,Creutz:2006wv,Creutz:2007yg}.
The required order is especially peculiar with one flavor
\cite{Creutz:2006wv}.}

For two or more flavors, this criticism is misguided.
Computers have a finite memory, so one takes the continuum limit,
$a\to0$, at fixed spatial volume, $L^3$.
But there is no spontaneous symmetry breaking in a finite volume.
To select a finite-volume vacuum close to the infinite-volume 
spontaneously-broken vacuum, explicit symmetry breaking is needed.
For this general reason, one should keep $m\neq0$ while taking
the continuum limit, then take $L\to\infty$, and last $m\to0$.

Also, the continuum limit must be carefully specified whenever exact
symmetries imply pseudo-Goldstone bosons~\cite{Hasenfratz:1992pngb}.
Suppose there are two kinds of particles with masses
$m_\pi^2a^2=\kappa\Sigma$, $m_\sigma^2a^2=\Sigma^2$, and consider an
unconventional family of continuum limits, $\kappa=m_qa^{1+p}$, 
$\Sigma=\Lambda a^{1-p}$.
Then $m_\pi^2=m_q\Lambda$, $m_\sigma^2=\Lambda^2a^{-2p}$, so if $p>0$
then $\pi$'s correlation length diverges while $\sigma$'s does not.
This is a different universality class than the standard one, $p=0$.
Staggered fermions (without rooting) are subtler still, because the
would-be $\pi$-like particles, except the one with taste~$P$, have a
mass $m_\xi^2a^2=\kappa\Sigma+\Sigma^4$,
$m_\xi^2=m_q\Lambda+\Lambda^4a^{2-4p}$.
For $p>1/2$, the continuum limit strands these at the cutoff,
reminiscent of the $\epsilon'$~regime.

With one flavor there is no spontaneous symmetry breaking---the  
lowest-lying pseudoscalar is an $\eta'$-like meson.
In this case, the above considerations no longer apply.
But every numerical lattice QCD calculation has unphysically large quark
masses, to allow the algorithms to run faster, so I do not consider
the order of limits to be a serious criticism even with one flavor.

\subsection{Mutilated quark-mass dependence}
\label{sub:mutilation}
\noindent\textsl{Rooted staggered fermions yield the same system for 
$-m$ as for $m$; therefore, the small-mass behavior is a function
of $m^2$, which we know is wrong in specific
cases~\cite{Creutz:2006ys,Creutz:2006wv,Creutz:2007yg}.}

\nopagebreak
Odd powers of the quark mass stem from zero modes of $\Dirac$.
As is obvious from Eq.~(\ref{eq:cleverBoy}), the rooting procedure 
turns~$m$ into $(m^4)^{1/4}=|m|$.
This defect of the algorithm has nothing to do with staggered 
fermions~\cite{Bernard:2006vv}.
\pagebreak

It is instructive to examine how odd powers arise when one has 
near-zero modes $\lambda_i\sim\Lambda^2a$, as with 
staggered fermions.
Two pairs with eigenvalue $\pm ic_i\Lambda^2a$, $i=1,2$
arise~\cite{Follana:2004sz}, so
\vspace*{-1pt}
\begin{equation}
	\left[{\det}_4(\Dirac_{\mathrm{stag}}+m)\right]^{1/4} \propto \left[
		\left(c_1^2\Lambda^4a^2 + m^2\right)
		\left(c_2^2\Lambda^4a^2 + m^2\right) \right]^{1/4} \approx 
		|m|\left[1 + 
		\case{1}{4}(c_1^2+c_2^2)\frac{\Lambda^4a^2}{m^2} \right]
	\label{eq:mutilated}
\vspace*{-1pt}
\end{equation}
when the limits are taken in the correct order.
If one takes the limits in the wrong order, then the determinant is
indeed a function of $m^2$.
It is correct to consider this the wrong theory, but it is incorrect to
assert that the numerical work takes the limit in the wrong order.
The applicability of Eq.~(\ref{eq:mutilated}) has been demonstrated 
clearly in the Schwinger model~\cite{Durr:2003xs}.
It is also possible to follow the $m$ dependence of observables down
until $m\sim\Lambda^3a^2\ll\Lambda^2a$ using 
RS$\chi$PT~\cite{Bernard:2004ab}, well beyond the regime where the 
approximation in Eq.~(\ref{eq:mutilated}) applies.

\subsection{Cancellation of non-unitary contributions}
\label{sub:cancellation}

\noindent\textsl{The cancellations among taste multiplets seem
contrived~\cite{Creutz:2006ys,Creutz:2006wv,Creutz:2007yg}.}

The \emph{Gedanken} algorithm of sect.~\ref{sec:rooting} shows that
phantoms cancel each other automatically in the absence of 
taste-exchange processes~\cite{Bernard:2006vv}.
Thus, if the Shamir defect scales so that one can treat it as a 
perturbation, it is not reasonable to call the cancellation contrived.
The scalar propagator provides a good numerical test, for two reasons.
First, the weights of the non-unitary contributions depend (in
$\chi$PT) on $N_r$ in an illuminating way, which could be checked 
numerically with (some) new simulations with $N_r=1,2,3,4$.
Second, the cancellation of the phantom modes depends on~$a^2$, which 
is being monitored as part of ongoing simulations~\cite{Bernard:2007qf}.

\subsection{Rank of chiral symmetry}
\label{sub:rank_of_chiral_symmetry}

\noindent\textsl{The rank (\emph{i.e.}, number of diagonal generators) 
of the chiral symmetry is wrong~\cite{Creutz:2006wv,Creutz:2007yg}.}

With four tastes the rank is indeed larger.
But the \emph{Gedanken} algorithm shows that this is not a feature of 
staggered fermions per se, but of the rooting procedure itself.
It provides a clear explanation of why the \emph{physical} sector is the
taste-singlet sector.
The other particles are phantoms.
In particular the extra neutral pseudoscalars cancel each other out,
perfectly in the \emph{Gedanken} algorithm and (on the basis of
numerical results) to order $a^2$ with rooted staggered fermions.

\subsection{Anomalies}
\label{sub:anomaly}

\noindent\textsl{The conventional axial anomaly 
cancels~\cite{Creutz:2006ys,Creutz:2006wv,Creutz:2007yg}.}

This claim is simply wrong.
The taste-$P$ axial current suffers no anomaly, as desired, because it
is off-diagonal in taste.
The conventional \emph{staggered} anomaly appears in the taste-singlet 
PCAC relation, as explained in the discussion of\
Eqs.~(\ref{eq:current}) and~(\ref{eq:pseudoscalar}).
It comes with the right strength, because rooting multiplies the
4-species anomaly with the factor $\case{1}{4}$ appropriate for one
species.

\subsection{Topology}
\label{sub:topology}

\noindent\textsl{Rooted staggered quarks implement topology incorrectly, 
because quartets must break up at the boundary of topological 
sectors~\cite{Creutz:2006ys,Creutz:2007yg,Creutz:2007nv}.
Rooting averages over positive and negative chirality modes, so the
index theorem cannot be satisfied~\cite{Creutz:2006wv}.}

Studies of the eigenvalues reveal a quartet structure, including 
quartets of near-zero modes~\cite{Follana:2004sz}.
As the gauge field transits \pagebreak from one topological sector to another, this
structure must indeed be disrupted.
But gauge fields near the boundary of topological sectors should have a
large gluon action and are, hence, suppressed.
If not, it is a drawback of the gluon action and has nothing to do 
with staggered fermions.
It would be interesting to monitor this with smooth gauge fields.
It would bode well for rooting if the quartets rearrange themselves 
rapidly, similarly to the way a lattice approximant to $Q$~does.

The assertion about positive and negative chirality is false.
Ref.~\cite{Wong:2004nk} shows that near-zero modes appear in quartets
of the same taste-\emph{singlet} chirality, Eq.~(\ref{eq:pseudoscalar}).
The confusion may stem, as in sect.~\ref{sub:anomaly}, from 
contemplating taste-$P$ currents and densities.
But taste-$P$ chirality vanishes for \emph{all} modes, including 
near-zero modes: $\sum_x f_i^\dagger(x)\varepsilon(x)f_i(x)=0$, because
if $f_i(x)$ is an eigenvector of eigenvalue $i\lambda_i+m$, then
$\varepsilon(x)f_i(x)$ is the eigenvector of $-i\lambda_i+m$, and they 
are orthogonal.
Hence, many statements about chirality in Refs.~\Creutz\ are simply
ill-conceived.

\subsection{'t~Hooft vertices}
\label{sub:t_hooft_vertices}

\noindent\textsl{'t~Hooft vertices generate contributions to correlation
functions that diverge in the chiral limit as a power of
$m^{-1}$~\cite{Creutz:2007nv,Creutz:2007rk}.}

Only a full analysis of 't~Hooft vertices can refute this assertion,
because no explicit equations are given in
Refs.~\cite{Creutz:2007nv,Creutz:2007rk}.
Here I shall examine the $\eta'$-like meson with one quark flavor, 
which should~\cite{Creutz:2007nv,Creutz:2007rk} expose the problem.
The taste-singlet propagator consists of two terms
\begin{eqnarray}
	C(x,y) & = & \tr[G(x,y)U_5      G(y,x)U_5] ,
	\label{eq:connected-eta-prime}  \\
	D(x,y) & = & \tr[G(x,x)U_5] \tr[G(y,y)U_5] ,
	\label{eq:disconnected-eta-prime}
\end{eqnarray}
where the trace is over color, and $U_5$ abbreviates the sign
factors, link matrices, and translations indicated in
Eq.~(\ref{eq:pseudoscalar}).
The quark propagator 
$G(x,y) = \langle\chi(x)\bar{\chi}(y)\rangle_{\chi,\bar{\chi}}$.
The right way to combine $C(x,y)$ and $D(x,y)$ is
\begin{equation}
	\left\langle\eta'(x)\eta'(y)\right\rangle = \left\langle
		-\case{1}{4} C(x,y)+\case{1}{16}D(x,y) \right\rangle_U,
	\label{eq:all-eta-prime}
\end{equation}
where the sign arises from Fermi statistics.
The weights $\case{1}{4}$ and $\case{1}{16}$ are crucial and follow 
immediately from Eqs.~(\ref{eq:originalZ}) and~(\ref{eq:cleverZ}).

't~Hooft vertices arise from (near) zero modes.
At first glance, such gauge fields are suppressed by small eigenvalues 
from the determinant.
For rooted staggered fermions the determinant factor is that given in
Eq.~(\ref{eq:mutilated}).
In a correlation function, however, these factors can be cancelled by
small eigenvalues in the denominator, coming from fermion propagators.

Let us consider the $Q=1$ sector for simplicity.
Numerical simulation~\cite{Follana:2004sz} tells us
there are four near-zero modes, which we shall label $\pm1$, $\pm2$.
Inserting an eigenvector-eigenvalue representation of the fermion
propagators, the disconnected contribution is
\begin{equation}
	D(x,y) = \sum_{i=\pm1,\pm2}
		\frac{1}{i\lambda_i+m} f^\dagger_i(x)U_5f_i(x) 
		\sum_{j=\pm1,\pm2}
		\frac{1}{i\lambda_j+m} f^\dagger_j(y)U_5f_j(y) + \ldots
		\sim \left(\frac{4}{m}\right)^2,
	\label{eq:disconnected-zero-mode}
\end{equation}
isolating the most singular parts as $m\to0$ (with $\lambda\ll m$).
Similarly, the connected correlator is
\begin{equation}
	C(x,y) = \sum_{i,j}
		\frac{1}{i\lambda_i+m} f^\dagger_i(x)U_5f_j(x)
		\frac{1}{i\lambda_j+m} f^\dagger_j(y)U_5f_i(y).
\end{equation}
Within a quartet, it is plausible to assume
$f^\dagger_i(x)U_5f_j(x)\sim{\rm O}(a)$, $i\neq j$, because each
eigenvector should have a different taste.  \pagebreak
Then,
\begin{equation}
	C(x,y) = \sum_{i=\pm1,\pm2}
		\left(\frac{1}{i\lambda_i+m} \right)^2
		f^\dagger_i(x)U_5f_i(x) f^\dagger_i(y)U_5f_i(y)  + \ldots 
		\sim \frac{4}{m^2}.
	\label{eq:connected-zero-mode}
\end{equation}
When combining the two pieces according to Eq.~(\ref{eq:all-eta-prime}),
no singular behavior appears as $m\to0$ (after $a\to0$).
Thus, modulo one easy-to-check assumption, 
Creutz's claims about the 't~Hooft vertex do not hold up in this, the 
simplest, example.
Similarly, the 't~Hooft vertex of the \emph{physical} eight-fermion
operator discussed in Ref.~\cite{Creutz:2007rk} is not singular, when 
Fermi statistics and orthogonality in taste is taken into 
account~\cite{Bernard:2007gss}.

\subsection{Summary}
\label{sub:summary_refutation}

The criticisms of Refs.~\Creutz\ are based on three sources of confusion.
Difficulties with mass dependence, the rank of the flavor-taste
symmetry, and the way that phantoms cancel are explained by the
\emph{Gedanken} algorithm of sect.~\ref{sec:rooting}.
This approach also illuminates the blunder of confusing the exact 
taste-nonsinglet chirality with the conventional taste-singlet 
chirality, which connects correctly to topology and 't~Hooft vertices.
Finally, it is easy to draw incorrect conclusions by choosing the incorrect 
order of limits (obdurately forcing $m\to 0$ before $a\to 0$).
The correct order is not ``absurd'' \cite{Creutz:2006ys,Creutz:2007yg}, 
but necessary whether or not staggered fermions are employed.

\section{New developments}
\label{sec:new_developments}

There are two noteworthy methodological developments in improved 
actions for staggered fermions.
One is a new discretization reducing taste-exchange interactions, while
maintaining O($a^2$) improvement.
The other is the completion of the full O($\alpha_sa^2$) corrections to
the gluon action.
Either or both could be incorporated into future simulations of sea 
quarks.

\subsection{Highly improved staggered quarks}
\label{sub:HISQ}

Although the Asqtad action has much smaller taste-changing effects than 
the original staggered action, it would be better to reduce them 
further.
The Asqtad action is obtained from the standard staggered action, 
Eq.~(\ref{eq:staggered}), in two steps, as follows.
First smeared links are constructed,
\begin{equation}
	V_\mu = \mathcal{F}_\mu U_\mu =
		\prod_{\rho\neq\mu}^{\rm sym} \left(1 +
		\frac{a^2}{4}\triangle_\rho \right) U_\mu,
\end{equation}
where $\triangle_\rho$ is a covariant nearest-neighbor second
derivative, and the product is symmetrized over all possible orderings
of the directions orthogonal to~$\mu$.
It yields bent staples of length 3, 5, and~7; substituting $V$ for $U$ 
in Eq.~(\ref{eq:staggered}) yields the FAT7 action.
The smearing is designed for, and is successful at, reducing the size 
of taste-exchange interactions.
It does not achieve Symanzik improvement, however.
This is achieved by adding two improvement terms to the FAT7 action, the
Naik~\cite{Naik:1986bn} term and the Lepage term~\cite{Lepage:1998vj},
to obtain the $a$-squared (Asq) action and, with tadpole-improved 
couplings, the Asqtad action.

A new action~\cite{Follana:2006rc} extends the philosophy behind the
FAT7 and Asqtad actions, introducing
$W_\mu = \mathcal{F}_\mu\mathcal{U}\mathcal{F}_\mu U_\mu$,
where the operator $\mathcal{U}$ brings the smeared $V$-links back into
U(3), thereby reducing ultraviolet fluctuations, including taste-changing
interactions~\cite{Hasenfratz:2001hp}.
Using $W$ in Eq.~(\ref{eq:staggered}) and generalizing
the Naik and Lepage terms yields the highly-improved
staggered quark (HISQ) action~\cite{Follana:2006rc}.
A numerical implementation~\cite{Wong:2007uz} is only around two times
slower than Asqtad. \pagebreak

The first results from the HISQ action shed light on the issues of
sect.~\ref{sec:synthesis}.
Reference~\cite{Follana:2006rc} calculates in perturbation theory the
size of the four-quark interactions in the Symanzik LE$\mathcal{L}$ for
both Asqtad and HISQ, finding the latter's to be an order of magnitude
smaller.
Reference~\cite{Follana:2006rc} also computes the $a^2\Delta_\xi$
splittings of Eq.~(\ref{eq:mass2}), finding them to be half as large
with HISQ.
In the language of (\ref{eq:BGS}) and~(\ref{eq:source}), the HISQ action
reduces the size of the (valence) Shamir defect, $va\Delta_n$, by
methodological means, and yields a valence spectrum closer to that
of the continuum limit.

\subsection{$\bm{n_f}$ dependence of the gauge action}
\label{sub:_n_f_dependence_of_the_gauge_action}

The MILC ensembles~\cite{Bernard:2001av} use the order-$a^2$ improved
gauge action~\cite{Weisz:1982zw}.
The pure-gauge one-loop matching of this action has been available for a
long time~\cite{Luscher:1985zq}.
Using these improved couplings, as MILC does, removes errors formally of
order~$\alpha_sN_ca^2$ but not~$\alpha_sn_fa^2$.
Earlier this year, Hao \emph{et al.}\ completed the fermion loop
calculation with Asqtad fermions~\cite{Hao:2007iz}.
They find that the fermion loop has the opposite sign from the
gluon+ghost loops.
For $n_f=3$ they change the sign of the radiative correction to the
coupling of the rectangle and reduce greatly the radiative correction to
the twisted parallelogram.
They recommend putting these results into future simulations and note 
that the size of the effects is what is needed to explain observed 
scaling violations~\cite{HartHippel}.
The MILC Collaboration plans to use this result in future 
simulations~\cite{Toussaint}.

\section{Conclusions}
\label{sec:conclusions}

Staggered fermions are fast, but not easy.
After trivializing the spinor index of the na\"ive fermion field, the
projection to one component entangles flavor symmetries with spacetime
symmetries, and the remaining species doubling is reflected in a new
flavor-like quantum number, taste.
The exact flavor and taste symmetries are expected to enlarge to 
(softly broken) $\mathrm{SU_V}(4n_f)\times\mathrm{SU_A}(4n_f)$, a 
mechanism that has been established by numerical simulations.

When the fourth root is taken, staggered fermions have both too little 
symmetry (discretization effects break 
$\mathrm{SU_V}(4n_f)\times\mathrm{SU_A}(4n_f)$) and too much
(the target symmetry is $\mathrm{SU_V}(n_f)\times\mathrm{SU_A}(n_f)$).
Confusion stemming from too much symmetry can be avoided via the
\emph{Gedanken} algorithm discussed in sect.~\ref{sec:rooting}.
It explains why rooted theories have ``extra'' particles, in
particular pseudo-Goldstone bosons, and why the extra symmetry protects
physical (single-taste) correlators.
Most of the criticisms of Refs.~\Creutz\ can be refuted with 
this framework~\cite{Bernard:2006vv,Bernard:2007gss}.

Valid criticisms of rooted staggered fermions should focus on the 
difference between staggered fermions and the continuum, namely on the 
interactions that break the full
$\mathrm{SU_V}(4n_f)\times\mathrm{SU_A}(4n_f)$ taste-flavor symmetry.
The most glaring issue is the violation of unitarity.
With the full symmetry such violations cancel identically (for physical 
correlators).
Without the full symmetry they do not cancel, but rooted staggered
chiral perturbation theory (RS$\chi$PT) offers a way to describe them.
\nopagebreak
Fits to pseudoscalar masses and decay constants (light and heavy-light),
and studies of the two-particle contribution to the scalar propagator,
give evidence that RS$\chi$PT works.
It not only guides the chiral extrapolation but also provides a
framework for estimating the associated uncertainty.

Refutation of specific criticisms, here and in
Refs.~\cite{Bernard:2006vv,Bernard:2007gss}, does not prove that rooted,
staggered fermions are valid.
It is remarkable, however, that so many numerical tests have shored up
the theoretical framework, when any one of them could have gone wrong.
Of special interest here is the interplay of chirality and topology.
The analysis of chirality in Refs.~\Creutz\ is wrong, because it
discusses a taste-nonsinglet chiral symmetry that has nothing to do with
topology.
The conventional staggered taste-singlet chirality agrees with the 
index theorem, in the same way any two lattice definitions of 
topological charge do.

More tests can and should be carried out.
A comprehensive study of the scaling of the Shamir defect, $a\Delta_n$, 
should help decide whether all taste breaking interactions vanish in the
continuum limit.
For self-consistency, this study should be done with a rooted staggered 
sea.
Tests of the sea, such as the scaling of the two-particle contributions
to taste-singlet correlators, would bolster confidence in RS$\chi$PT.
Most of Creutz's discussion of 't~Hooft vertices can be refuted with 
the setup in sect.~\ref{sec:rooting}, but it does bring out the need to 
check numerically whether the members of near-zero mode quartets are 
all of different taste.

Much of our structural, and not to mention practical, understanding of
rooted, staggered fermions depends on RS$\chi$PT.
It is based on plausible arguments~\cite{Bernard:2006zw}, 
with one reservation.
An original justification for $\chi$PT is that it gives the most general
description of particle interactions, consistent with
unitarity~\cite{Weinberg:1978kz}.
What happens when unitarity is lost?
Can a non-unitary $\chi$PT describe a non-unitary gauge theory?
Is, for example, cluster decomposition enough~\cite{Bernard:2007ycd}?
This is a basic concern wherever partially quenched $\chi$PT is used.
With other formulations of lattice fermions one could, if necessary,
avoid it.
With staggered fermions it seems essential.

A friend of mine, who expects that Eq.~(\ref{eq:rooting}) is not valid,
says that the burden of proof is on the staggered community.
He is correct, of course, but only half correct.
A ``proof'' is unlikely to be mathematically rigorous.
Instead, as in all of lattice gauge theory, methods will be validated, 
or not, with a combination of theoretical framework and numerical 
simulation.
It does not make much sense for proponents to prove to themselves that 
their methods are acceptable.
Skeptics need to be engaged, examine the theoretical and numerical
evidence in favor of rooted staggered fermions, think about the issues
clearly, and state where the shortcomings lie.
In other words, their job now is to be like Mike.
Think it through and write it up!

\acknowledgments
I would like to thank Mike Creutz for agreeing to the format of our
talks and, most of all, for documenting his criticisms in the
literature, where they can be scrutinized openly.
While preparing the oral and written versions of this talk, I benefited
from conversations and e-mail correspondence with
David Adams,
Christopher Aubin,
Jon Bailey,
Bill Bardeen,
Claude Bernard, 
Poul Damgaard,
Christine Davies,
Carleton DeTar,
Stefan D\"urr,
Eduardo Follana,
Elizabeth Freeland,
Maarten Golterman, 
Alistair Hart,
Peter Hasenfratz,
Georg von~Hippel,
Jack Laiho,
Yigal Shamir,
Steve Sharpe, 
Doug Toussaint, 
and \nopagebreak
Ruth Van de Water.

\end{document}